\documentclass{IEEEtran4PSCC}

\ifCLASSINFOpdf
   \usepackage[pdftex]{graphicx}
\else
   \usepackage[dvips]{graphicx}
\fi

\usepackage[cmex10]{amsmath}

\usepackage{color}

\usepackage[colorlinks=true, allcolors=blue]{hyperref}
\usepackage[colorinlistoftodos]{todonotes}
\usepackage{graphicx}% Include figure files
\usepackage{xcolor}
\usepackage{dcolumn}% Align table columns on decimal point
\usepackage{bm}% bold math
\usepackage{silence}
\usepackage{relsize}
\usepackage{comment}
\usepackage{subcaption} % Add the subcaption package
\usepackage{svg}
\usepackage[english]{babel}
\makeatletter

\let\old@ps@headings\ps@headings
\let\old@ps@IEEEtitlepagestyle\ps@IEEEtitlepagestyle
\def\psccfooter#1{%
    \def\ps@headings{%
        \old@ps@headings%
        \def\@oddfoot{\strut\hfill#1\hfill\strut}%
        \def\@evenfoot{\strut\hfill#1\hfill\strut}%
    }%
    \def\ps@IEEEtitlepagestyle{%
        \old@ps@IEEEtitlepagestyle%
        \def\@oddfoot{\strut\hfill#1\hfill\strut}%
        \def\@evenfoot{\strut\hfill#1\hfill\strut}%
    }%
    \ps@headings%
}

\newcommand{\linebreakand}{%
  \end{@IEEEauthorhalign}
  \hfill\mbox{}\par
  \mbox{}\hfill\begin{@IEEEauthorhalign}
}

\makeatother

\psccfooter{%
        \parbox{\textwidth}{\hrulefill \\ \small{23rd Power Systems Computation Conference} \hfill \begin{minipage}{0.2\textwidth}\centering \vspace*{4pt} \includegraphics[scale=0.06]{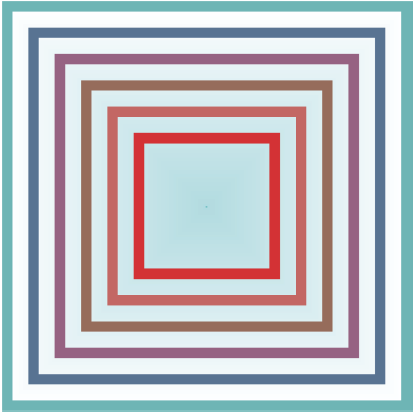}\\\small{PSCC 2024} \end{minipage} \hfill \small{Paris, France --- June 4 -- June 7, 2024}}%
}

\IEEEoverridecommandlockouts
\IEEEaftertitletext{\vspace{-1.5\baselineskip}}

\begin{document}

\title{Non-standard power grid frequency statistics in Asia, Australia, and Europe}

\author{
\IEEEauthorblockN{\textbf{1\textsuperscript{st} Xinyi Wen}}
\IEEEauthorblockA{\textit{Institute for Automation and} \\ \textit{Applied Informatics (IAI)} \\ \textit{Karlsruhe Institute of Technology (KIT)} \\
Eggenstein-Leopoldshafen, Germany \\
\ xinyi.wen@kit.edu}

\and
\IEEEauthorblockN{\textbf{2\textsuperscript{nd} Mehrnaz Anvari}}
\IEEEauthorblockA{\textit{Fraunhofer-Institute for Algorithms} \\ \textit{and Scientific Computing (SCAI)}\\
Sankt Augustin, Germany\\
\ mehrnaz.anvari@scai.fraunhofer.de}

\and
\IEEEauthorblockN{\textbf{3\textsuperscript{rd} Leonardo Rydin Gorjão}}
\IEEEauthorblockA{\textit{Faculty of Science and Technology,} \\
\textit{Norwegian University of Life Sciences}\\
Ås, Norway\\
\ leonardo.rydin.gorjao@nmbu.no
\linebreakand
%\and
\IEEEauthorblockN{\textbf{4\textsuperscript{th} G.Cigdem Yalcin}}
\IEEEauthorblockA{\textit{Department of Physics,} \\
\textit{Istanbul University}\\
Istanbul, Turkey\\
\ gcyalcin@istanbul.edu.tr}
}
\and
\IEEEauthorblockN{\textbf{5\textsuperscript{th} Veit Hagenmeyer}}
\IEEEauthorblockA{\textit{Institute for Automation and} \\ \textit{Applied Informatics (IAI)} \\
\textit{Karlsruhe Institute of Technology (KIT)}\\
Eggenstein-Leopoldshafen, Germany\\
\ veit.hagenmeyer@kit.edu}

\and
\IEEEauthorblockN{\textbf{6\textsuperscript{th} Benjamin Schäfer}}
\IEEEauthorblockA{\textit{Institute for Automation and} \\ \textit{Applied Informatics (IAI)} \\
\textit{Karlsruhe Institute of Technology (KIT)}\\
Eggenstein-Leopoldshafen, Germany\\
\ benjamin.schaefer@kit.edu}
}

% Make the title area
\maketitle

\begin{abstract}
The power-grid frequency reflects the balance between electricity supply and demand. Measuring the frequency and its variations allows monitoring of the power balance in the system and, thus, the grid stability. 
In addition, gaining insight into the characteristics of frequency variations and defining precise evaluation metrics for these variations enables accurate assessment of the performance of forecasts and synthetic models of the power-grid frequency.
Previous work was limited to a few geographical regions and did not quantify the observed effects. 
In this contribution, we analyze and quantify the statistical and stochastic properties of self-recorded power-grid frequency data from various synchronous areas in Asia, Australia, and Europe at a resolution of one second. Revealing non-standard statistics of both empirical and synthetic frequency data, we effectively constrain the space of possible (stochastic) power-grid frequency models and share a range of analysis tools to benchmark any model or characterize empirical data. Furthermore, we emphasize the need to analyze data from a large range of synchronous areas to obtain generally applicable models.
\end{abstract}

\begin{IEEEkeywords}
bimodal, frequency, linear test, correlation, SDE modeling, power grid, Hurst exponent, statistics, heavy tails \vspace*{-1.5em}
\end{IEEEkeywords}

% Use this to place sponsorships
\thanksto{\noindent Submitted to the 23nd Power Systems Computation Conference (PSCC 2024).}

\section{Introduction}

% What can the power grid do
A power grid is a complex and interconnected network that enables the transmission and distribution of electricity from generators to consumers \cite{kaplan2009electric}.
It is a vital infrastructure ensuring homes, companies, and industries have access to a robust supply of electricity \cite{obama2013presidential}.
To operate, a power grid must maintain a constant balance between electricity supply and demand.
Any deviation from this balance can lead to grid instability, blackouts, or infrastructure damage. 
The power-grid frequency is a measurable quantity that indicates the operational status of a power grid. It reflects the rotational speed of the numerous synchronous machines within one area so that we refer to a region with one shared frequency as a synchronous area.
An excessive feed-in of power into the grid causes an increase in frequency, whereas an insufficient supply results in a decrease in frequency.
Sudden changes in this frequency can cause grid instability, which is why maintaining consistent frequency levels is important.
The power-grid frequency typically remains within a few percentage points of a reference value of 50\,Hz or 60\,Hz through the installation of various balancing and control systems in place \cite{kundur2022power}.
These control measures monitor and stabilize the frequency to keep it within a permissible range \cite{7936473,short2007stabilization,bevrani2014power}.

To use expensive control measures as efficiently as possible, e.g. via forecasting algorithms, a thorough understanding and modeling of the power-grid frequency is necessary.
Therefore, the analysis of the stochastic nature of the power-grid frequency has garnered significant interest and attention from mathematicians, statisticians, and physicists alike \cite{kundur2022power, rohden2012self, filatrella2008analysis}.
Non-Gaussian frequency distributions have been discussed for European synchronous areas \cite{schafer2018non, anvari2016short, 9744334}.
Furthermore, there have been studies that analyze frequency deviations and make predictions using the Fokker--Planck equation \cite{8626538}. 
However, a thorough characterization of the stochastic properties constraining potential models is missing. Moreover, previous works have often focused on European areas, while for example measurements from Asia have not been thoroughly investigated or compared.

We substantially expand previous research \cite{9744334, anvari2020stochastic} by conducting a rigorous quantitative analysis of the statistical properties of a large class of synchronous areas.
Our main objective is to establish quantitative measures that enable the comparison of different synchronous areas with each other and with synthetic models.
In particular, consider an equation of the form \begin{equation}
    \frac{df}{dt} = g(f, t) - \xi(f,t), \label{eq:genericModel}
\end{equation}
where $f$ is power-grid frequency, $g$ is the (unknown) intrinsic dynamics of the system, and  $\xi$ represents noise, e.g. $\xi=\frac{dW}{dt}$ where $W$ could be the Wiener process. Both $g$ and $\xi$ are potentially explicitly dependent on the frequency value $f$ and time $t$.
We now wish to understand how the empirical data constrains potential deterministic functions $g$ or stochastic contributions $\xi$.

This article is structured as follows: First, we give an overview of the multi-continent dataset (\ref{sec:data overview}).
We then investigate fundamental statistical properties and the frequency distribution of empirical power grid data.
Additionally, we compare the degree of bimodality across different synchronous areas (\ref{sec:Quantify bimodality}). 
Furthermore, we calculate the one-step increment in order to evaluate frequency fluctuations (\ref{sec:Log increment frequency}).
To establish a benchmark for comparison, we employ three distinct datasets derived from various stochastic differential models \cite{10202986}.
We then evaluate and compare the degree of linearity between empirical power grid data and synthetic data (\ref{sec:linearity}).
Lastly, we delve into a detailed analysis of the correlations in the system, suggesting non-Markovian behavior in the recorded frequency data (\ref{sec:markovian}). 
In the concluding section (\ref{sec:discussion}), we present our findings and engage in a comprehensive discussion to provide a deeper understanding of the observed statistical properties and dynamics of the power-grid frequency.  

\section{Data Overview}\label{sec:data overview}

Many previous studies, in particular ones discussing open data, have focused on European regions \cite{8626538,rydin2020open,jumar2021database,gorjao2022phase}. Meanwhile, there is limited research \cite{8973560} that systematically and quantitatively compares the frequency characteristics of power grids across Asia, Australia, and Europe.
One important reason is the limited availability of public data on power-grid frequency, making it difficult for researchers to comprehensively analyze the frequency behavior of power grids in different regions.
To address this issue, it is necessary to encourage data sharing and collaboration among industry actors and academics -- as well as to support initiatives that collect and disseminate such data.
Furthermore, conducting comparative analyses that consider these diverse geographical areas becomes essential. 
Indeed, by undertaking a systematic study of power-grid frequency characteristics across different geographical areas, we can examine the parallels, discrepancies, and underlying causes influencing power-grid frequency behavior. 

% How to get the data
To collect our dataset, we utilize a GPS-synchronized frequency acquisition device called an Electrical Data Recorder (EDR) developed at the Institute for Automation and Applied Informatics, Karlsruhe Institute of Technology, Germany \cite{maass2013first, maass2015data}. The EDR provides data similar to a Phasor Measurement Unit while allowing easy transfer and processing of the raw and processed signals.
Our primary mode of data collection involves connecting the EDR to conventional power sockets in an office or at a hotel to capture the voltage waveforms as experienced on the low-voltage distribution grid.
Such local voltage phasor measurements allow the extraction of frequency values,  which are essentially identical on the low-voltage and the high-voltage grid and thereby indicate the state of the entire synchronous area \cite{jumar2021database,6343992}.
Saving the original waveform information allows us to increase the accuracy of our data with further post-processing \cite{forstner2022experimental,schafer2023microscopic}.
In the present study, we collected power-grid recordings from various locations in the Southeast Asian region, including Indonesia, Malaysia, and Singapore, as well as measurements from the Australia National Electricity Market (NEM area). 

\begin{figure}
\centering
\includegraphics[width=1\columnwidth]{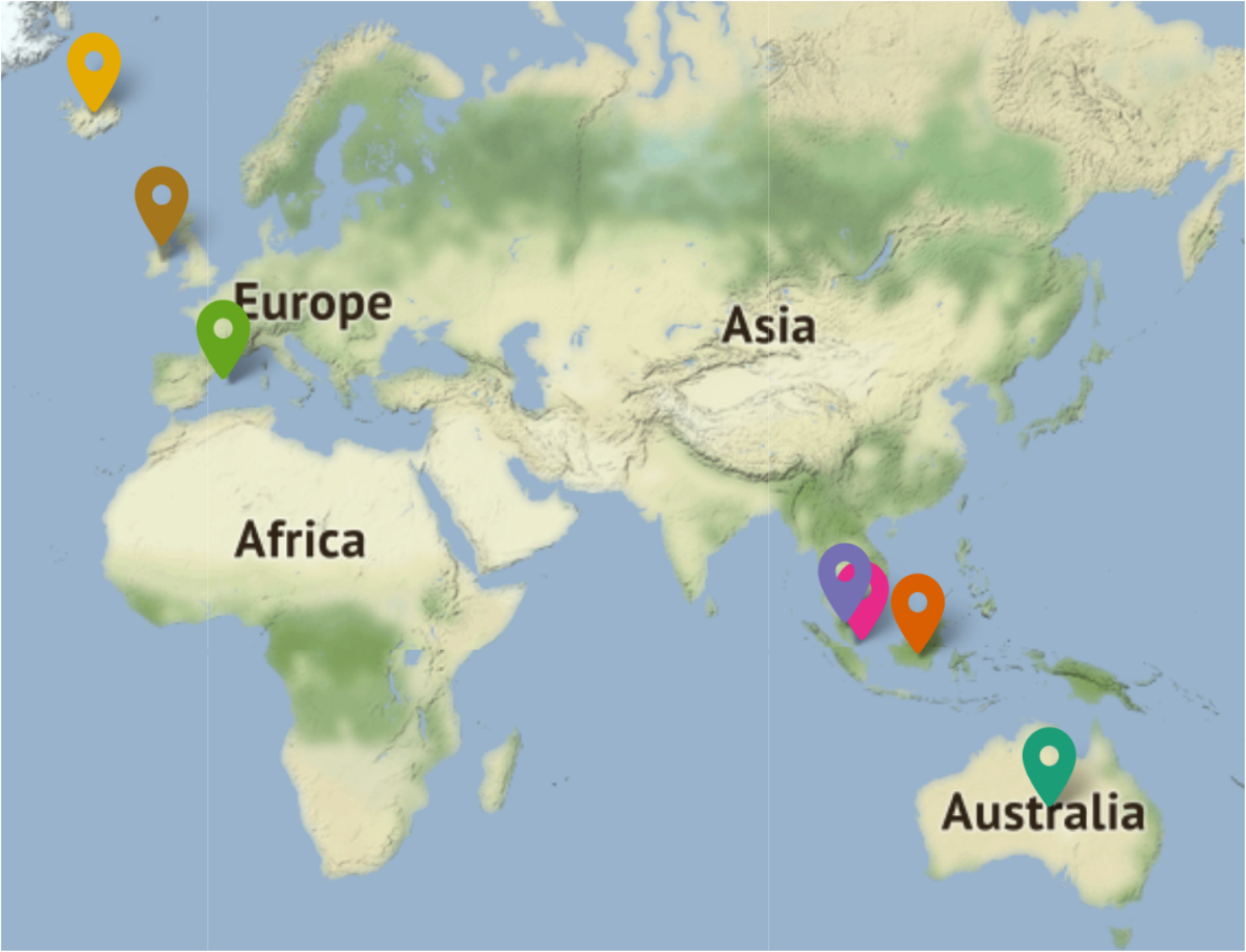}
\caption{\label{fig1} \textbf{Location overview}. Power-grid frequency recordings from Australia, Indonesia, Malaysia, Singapore, Iceland, Ireland, and the Balearic Islands. The map is created using the Folium package (OpenStreetMaps, Leaflet.js) in Python.}
\vspace*{-1.5em}
\end{figure}

% How many datasets, time range, and resolution
The data collection period at each location spanned 10 to 25 days, from October 30\textsuperscript{th}, 2022 to January 9\textsuperscript{th}, 2023. We evaluate the raw data to obtain frequency data with a resolution of one second.
The selection of these countries in this study is based on their geographical diversity and distinct characteristics in terms of generation mixture and grid configurations.
Each recording corresponds to a distinct synchronous region, wherein different countries exhibit a unique combination of energy sources, including fossil fuels and renewables.
This diversity in energy sources along with different operational control or market rules in each synchronous area contributes to variations in grid behavior and dynamics across these regions.
Furthermore, we also obtained power-grid frequency data from European areas, namely Iceland, Ireland, and the Balearic Islands.
For the European areas, the data collection timeframe was longer, ranging from September 29\textsuperscript{th}, 2019 to February 22\textsuperscript{nd}, 2022, covering at least three months in each location.
In Fig.~\ref{fig1}, we highlight the location of the measurement points on a world map.
To perform an accurate analysis, we remove any intervals lacking EDR-recorded frequency data, ensuring a continuous dataset.

Additionally, we employ stochastic models, such as a Langevin process \cite{10202986} and a fractional Brownian motion-based model \cite{mandelbrot1968fractional} to generate synthetic data.
Incorporating these synthetically generated datasets in our analysis serves a dual purpose: firstly, it allows us to gain a more comprehensive understanding of the underlying processes, and secondly, it enables us to validate and verify the efficacy of the methods we have employed in our study.

\section{Quantify bimodality} \label{sec:Quantify bimodality}

% visualize kde plot
To gain an initial impression of the distribution of the power-grid frequency, we utilize kernel density estimation to display the probability density function (PDF).
This approach visualizes the shape of the distribution and identifies the central tendencies.
In addition, we compute the normalized third and fourth moment, the skewness, and the kurtosis of the data. The skewness is a measure of the asymmetry of a distribution, with a value of zero indicating a symmetrical (potentially normally) distributed dataset. Meanwhile, kurtosis measures the behavior of the tails, i.e. of large deviations. A Gaussian distribution has a kurtosis of 3 so values below 3 indicate light and above 3 indicate heavy tails. We consistently observe kurtosis values below 3 in contrast to earlier results for European data \cite{schafer2018non,anvari2020stochastic}.

\begin{figure}
  \includegraphics[width=\linewidth]{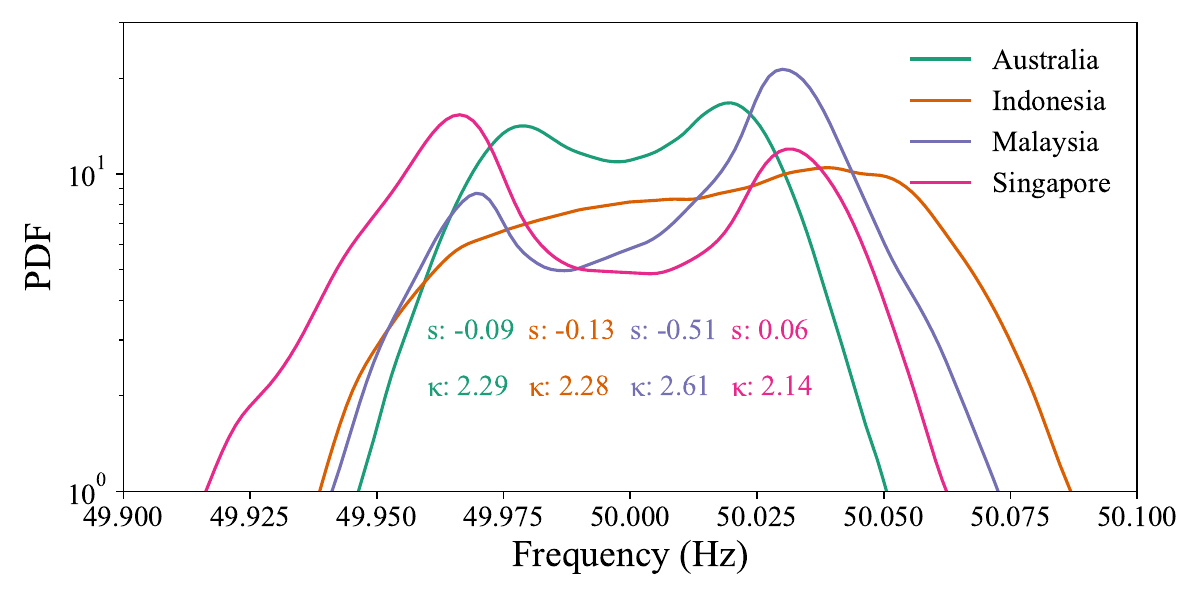}
  \caption{\label{fig2}\textbf{Frequency distribution in Asian areas}. All PDFs are shown on a vertical logarithmic scale. All power grids demonstrate non-Gaussian distributions, e.g. two peaks or non-Gaussian skewness $s$  and kurtosis $\kappa$ ($s^\text{Gauss}=0$, $\kappa^\text{Gauss}=3$).
}
\vspace*{-1.5em}
\end{figure}

All regions in our study operate at a reference value of 50\,Hz around which the grid frequency fluctuates.
Naively, we could expect that the most probable value of the grid frequency would be 50\,Hz.
Surprisingly, Fig.~\ref{fig2} shows that the PDF of the grid frequencies across Asian regions, with the exception of Indonesia, displays two peaks.
This indicates that the power-grid frequency follows a bimodal distribution, rather than a unimodal distribution, potentially due to deadbands in the control \cite{8626538,10202986}. 

\begin{figure}
\centering
\begin{subfigure}{0.5\textwidth}
  \centering
  \includegraphics[width=\linewidth]{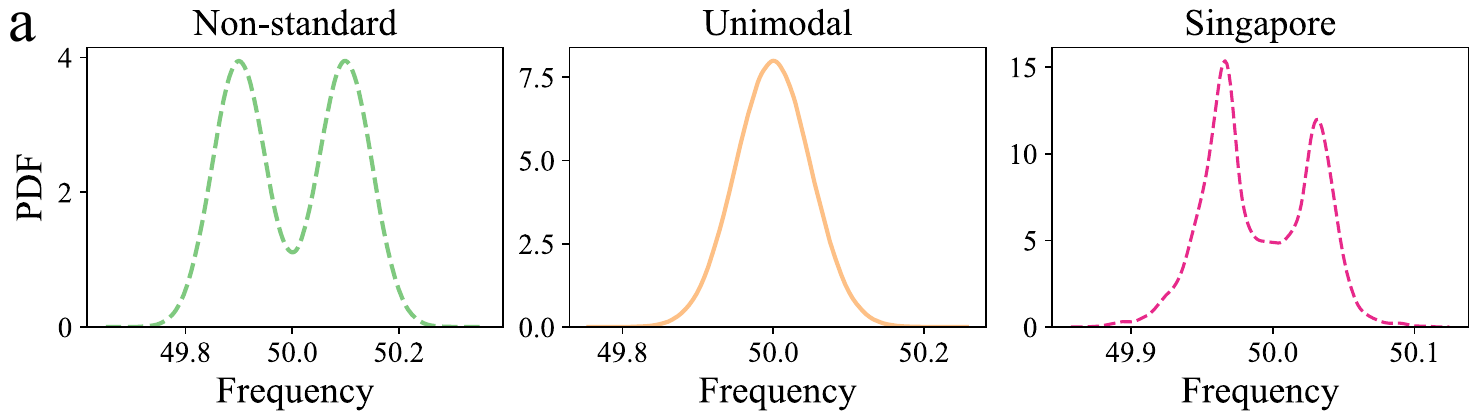}
\end{subfigure}
\begin{subfigure}{0.5\textwidth}
  \centering
  \includegraphics[width=\linewidth]{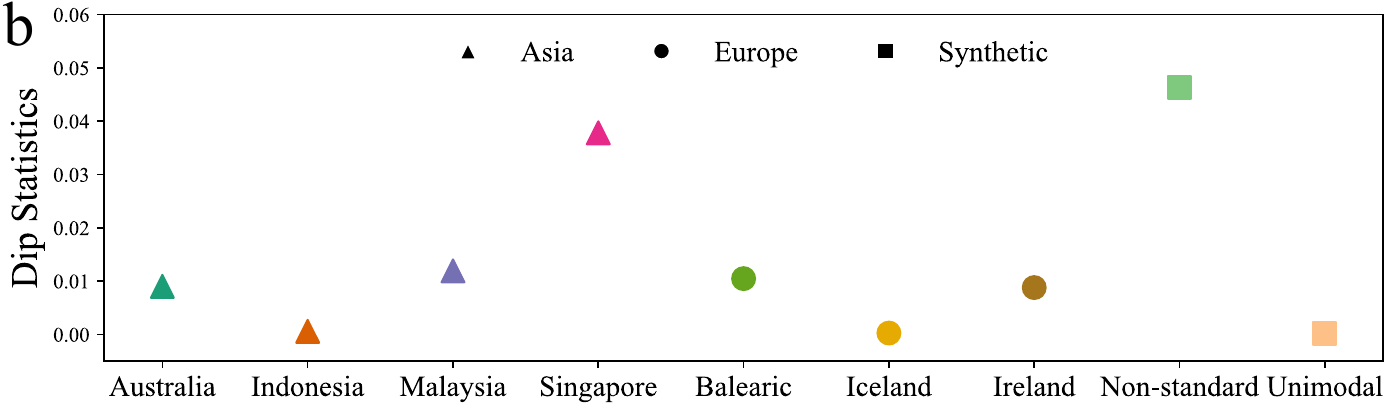}
\end{subfigure}
\captionsetup{justification=justified,singlelinecheck=false}
\caption{\textbf{Bimodality quantification}. \textbf{a}: The left plot shows synthetic data with a bimodal distribution, while the middle plot displays a normal distribution derived from synthetic data as well. The right plot, based on actual data, reveals a bimodal distribution. \textbf{b}: Using the dip statistics, we measure the level of bimodality, with Singapore showing the highest value.}
\vspace*{-1.5em}
\label{fig3}
\end{figure}

In order to illustrate the disparity in the distributional properties of our power-grid frequency data, Fig.~\ref{fig3}a showcases three distinct density curves.
The curves correspond to two synthetic baseline models and one empirical distribution observed in the data from Singapore.

% dip statistics
To quantify the distributional properties of our power-grid frequency data, we calculate the dip statistic \cite{hartigan1985dip}, which measures the degree of bimodality, or equivalently, the deviation from unimodality.
Specifically, it quantifies the distance between the empirical distribution and the closest unimodal distribution, with larger values indicating a greater departure from unimodality and providing more evidence for bimodality.
Fig.~\ref{fig3}b provides an overview of the dip statistic values for the power-grid frequency data collected in our study, as well as two synthetic datasets that follow a non-standard distribution and a unimodal distribution respectively, for reference. 
These plots allow us to compare the different distribution characteristics of the datasets.
Singapore demonstrates the highest degree of bimodality among the datasets examined, as indicated by the largest value of the dip statistic.
This result is consistent with our expectations, given the frequency distribution plot for Singapore shows the most pronounced double-peak pattern, see Fig.~\ref{fig2}.
Furthermore, our analysis indicates that Australia, Malaysia, the Balearic Islands, and Ireland also exhibit a bimodal distribution, evident from the relatively larger dip statistic values.

%Conversely, Australia, Indonesia, and Iceland show the least amount of deviation from unimodality, with the smallest values of dip statistic.
Conversely, both Indonesia and Iceland display zero values of dip statistics, implying a unimodal distribution.
This aligns with our expectations because their PDFs are more normally distributed compared to the other datasets \cite{rydin2020open}.

The observed bimodality indicates that the deterministic dynamics $g$ in \eqref{eq:genericModel} could originate from a double-welled potential or from a superposition of single-well statistics (superstatistics) \cite{schafer2018non}.

\section{Frequency increments}\label{sec:Log increment frequency}

% Why increment
To understand whether there exist sudden and extreme ramps or jumps in the frequency over time, we consider the distribution of the frequency increments for each region.

\begin{figure}
\centering
\begin{subfigure}{\columnwidth}
  \centering
  \includegraphics[width=\linewidth]{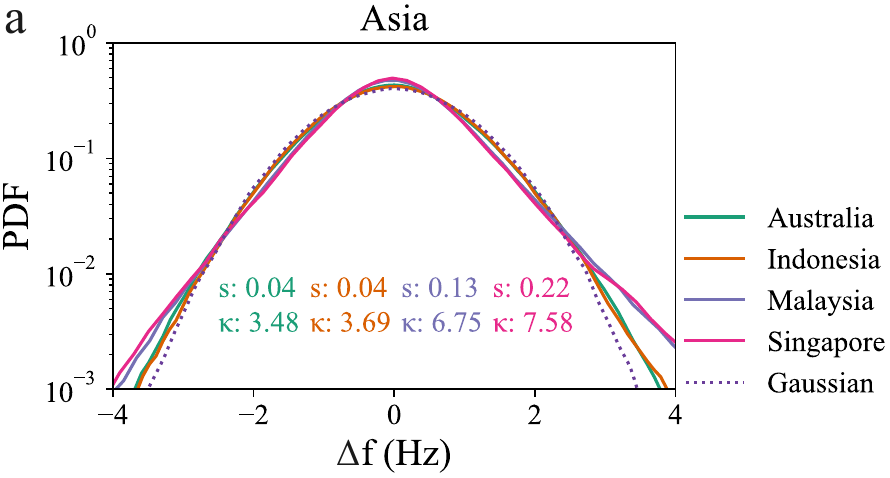}
  \label{fig4a}
\end{subfigure}
\vspace{-1\baselineskip}
\begin{subfigure}{\columnwidth}
  \centering
  \includegraphics[width=\linewidth]{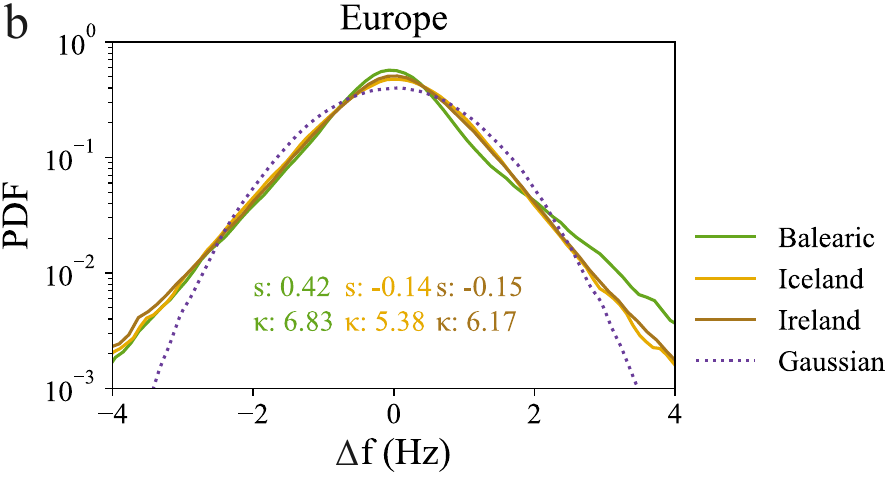}
  \label{fig4b}
\end{subfigure}

\captionsetup{justification=justified,singlelinecheck=false}
\caption{\textbf{Distribution of increment frequency}. Probability distributions of increment frequency on a vertical log scale. The variable $s$ represents skewness, and $\kappa$ represents kurtosis.}
\vspace*{-1.5em}
\label{fig4}
\end{figure}

% Increment frequency shows the unimodal (non-Gaussian) distribution
For this purpose, first, we calculate the frequency increment $\Delta f_{\tau}=f(t+\tau)-f(t)$, where $\tau=1$\,s, that is, the sampling rate of the data.
When we examine these increments, we observe that the PDF of the frequency increments $\Delta f = \Delta f_{\tau=1}$ of each region, as shown in Fig.~\ref{fig4}, exhibits deviations from a strict Gaussian distribution by displaying heavy tails for both negative and positive frequency increments.

% explain skewness, non-zero values of skewness
Additionally, to provide insights into the statistical properties and distribution characteristics of the frequency increments, we calculate the skewness and kurtosis of the increments for each region.
We observe that the Asian areas show positive skewness values, which can be understood as a ramp-up pattern, with a longer tail for positive values of the frequency increment distribution, see Fig.~\ref{fig4}a.
On the contrary, Iceland and Ireland exhibit negative skewness values, indicating a leftward skew with a longer tail for negative values of the frequency increment distribution. 
Still, our analysis reveals that all regions display a very small non-zero skewness, indicating deviations from symmetry are minor, see Fig.~\ref{fig4}.

% explain kurtosis, kurtosis greater than 3
Moreover, by measuring the kurtosis, we conclude that all synchronous areas exhibit a leptokurtic distribution with kurtosis values greater than 3. This suggests the presence of heavier tails in the frequency increment distribution compared to a Gaussian distribution. 

Our analysis indicates that the frequency increments in Asian and European areas do not follow a Gaussian distribution, as evidenced by the PDFs and statistical moments.
These findings align with prior research \cite{rydin2020open,anvari2020stochastic}.

The observed heavy tails and deviations from Gaussianity indicate that the stochastic dynamics $\xi$ in \eqref{eq:genericModel} might be better modeled as a Lévy-stable process or a superposition than a simple Wiener process \cite{schafer2018non}.

\section{Linearity}\label{sec:linearity}

% Why do we need to test linearity
In modeling power systems, there is often a preference for simplicity, and linear models are commonly employed.
This simplification might contrast with the nature of power systems themselves, where, e.g., power flow equations are nonlinear by nature, but the inertial response from generating units is linear.
However, it is unclear how this transpires to simplified models or empirical data of the power-grid frequency, and thus, we should test for the presence of linear contributions in the data.

% Brief describes the methods used to calculate the linearity 
We use the higher-order autocorrelation function to test the linearity of the power-grid frequency data by quantifying the correlation between two observations in a time series at a given time lag \cite{kantz2004nonlinear}.
The higher-order autocorrelation function of the frequency data provides a quantitative measure of time asymmetry in the dataset.
If a time series exhibits asymmetry in time, it is indicative of non-linearity in the underlying dynamics.
Therefore, we determine the degree of time asymmetry and quantify the level of non-linearity present in the power-grid frequency data.
We calculate the higher-order autocorrelation for a given data set as: 

\begin{equation}\label{eq:LT}
 LT (t) = \frac{[ f(t) - f(t + \tau) ]^3}{[ f(t) - f(t + \tau) ]^2},
\end{equation}
where $LT$ stands for "linear test".

To ensure the validity of our results for a realistic process, we compare the original data to a surrogate time series.
This involves taking the Fourier transform ($FT$) of the original data and randomizing the phases before using an inverse $FT$ to obtain the surrogate data.
The equation of Fourier transform ($FT$) is defined as:
\begin{equation}\label{eq:FT}
F(\omega) = \int\limits_{-\infty}^{\infty} f(t) \cdot e^{-i\omega t} \, dt, 
\end{equation}
and its inverse is given by
\begin{equation}\label{eq:inverse FT}
f(t) = \frac{1}{2\pi} \int\limits_{-\infty}^{\infty} F(\omega) \cdot e^{i\omega t} \, d\omega,
\end{equation}
where $\omega$ is the Fourier-frequency variable and $F(\omega)$ represents the Fourier transform of the function $f(t)$ (the power-grid frequency in our case).
The exponential term, $e^{-i\omega t}$, is the complex exponential function with an imaginary unit, $i$.
By implementing the procedure described above, we effectively eliminate any non-linearity in the process, resulting in surrogate data that solely reflects the linear characteristics of the analyzed data \cite{theiler1992testing}.

With the surrogate data at hand, we compute the $LT(t)$ for the empirical and the surrogate data and then quantify the distance between both time series using the root mean square error (RMSE).
If the value of the root mean square error (RMSE) is close to zero, it indicates that the time series exhibits linear behavior.

\begin{figure}
\centering
\includegraphics[width=1\columnwidth]{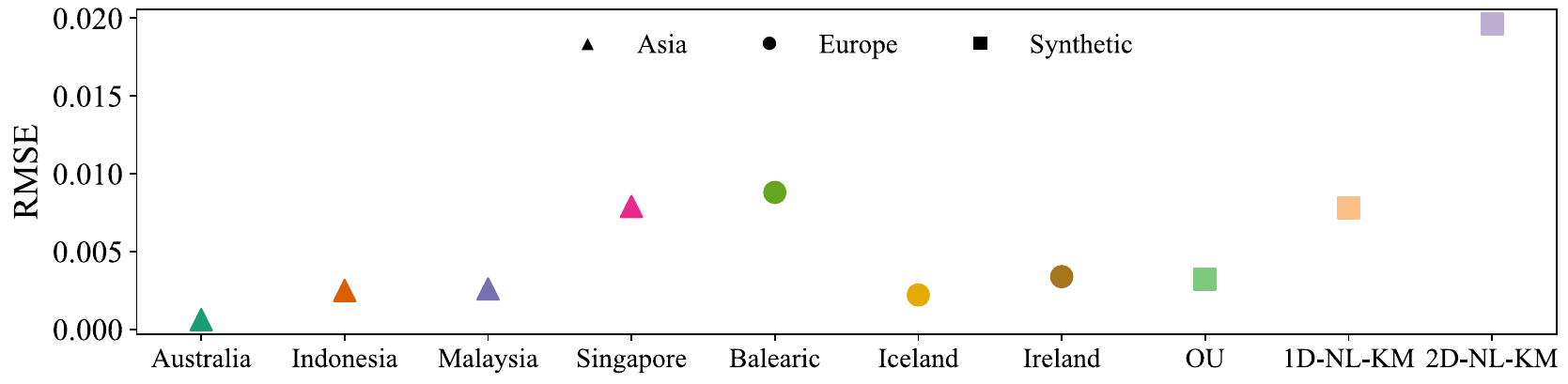}
\caption{\textbf{Linearity quantification}. Visualizing the degree of linearity $LT$ from \eqref{eq:LT}, measured by the RMSE of data vs. surrogate data. Lower values of RMSE indicate a smaller deviation from linearity.}
\vspace*{-1.5em}
\label{fig5}
\end{figure}

% looking at the plot and compare the rmse
To validate our findings and put them into context, we employ three synthetic datasets of Ireland provided by stochastic differential models \cite{10202986}.
For the sake of brevity, we will not detail the models extensively and point the reader to Oberhofer \textit{et al.}~\cite{10202986}.
We only include a short description of these stochastic models of the power-grid frequency, denoted as OU,1D-NL-KM, and 2D-NL-KM.
OU is a basic Ornstein--Uhlenbeck process with a single damping constant and noise term.
1D-NL-KM represents a one-dimensional non-linear Kramers–Moyal model that incorporates a non-linear response and multiplicative noise.
2D-NL-KM is a two-dimensional non-linear Kramers–Moyal model that separates the frequency into stochastic fluctuations and a deterministic trend.
Hence, OU is fully linear, while 1D-NL-KM and 2D-NL-KM are designed to include non-linear effects.
Our RMSE results attest to the nature of these models, as OU exhibits the lowest nonlinearity while 2D-NL-KM demonstrates the highest nonlinearity, as measured by their respective RMSE scores in Fig.~\ref{fig5}.
After analyzing the RMSE of the observed power-grid frequency data and three synthetic datasets, we find that the power-grid frequency data from Australia exhibits a linear property, characterized by small RMSE values of the LT test.
Meanwhile, the Singapore and Balearic regions exhibit the greatest RMSE values, indicating a larger deviation from linearity, see Fig.~\ref{fig5}.
However, even the largest RMSE value for the power-grid frequency data is significantly smaller than the non-linearity observed in the synthetic models.

The observed linearity indicates that the deterministic dynamics $g$ in \eqref{eq:genericModel} should approximately follow a linear relationship, which supports the idea of a superposition of simple statistics \cite{schafer2018non} over explicit nonlinear modeling \cite{8626538,10202986}.

\section{Correlation analysis}\label{sec:markovian}

% autocorrelation plot and decay constant
When modeling stochastic processes it is key to assess whether a process is Markov or not, i.e. whether there are long-range correlations present. Hence, we investigate the autocorrelation and decay characteristics of the datasets for Asian, Australian, and European power grids. 
The autocorrelation function (ACF) at lag $\tau$ of a time series \(f_t\) is calculated as:

\begin{equation}\label{eq:acf}
\text{ACF}(\tau) = \frac{\text{Cov}(f_t, f_{t-\tau})}{\sqrt{\text{Var}(f_t) \cdot \text{Var}(f_{t-\tau})}},
\end{equation}
where \(\text{Cov}(f_t, f_{t-\tau})\) is the covariance between \(f_t\) and \(f_{t-\tau}\), \(\text{Var}(f_t)\) is the variance of the original series at time \(t\), and \(\text{Var}(f_{t-\tau})\) is the variance of the lagged series at time \(t-\tau\).

\begin{figure}
\centering
\begin{subfigure}{\columnwidth}
  \centering
  \includegraphics[width=\linewidth]{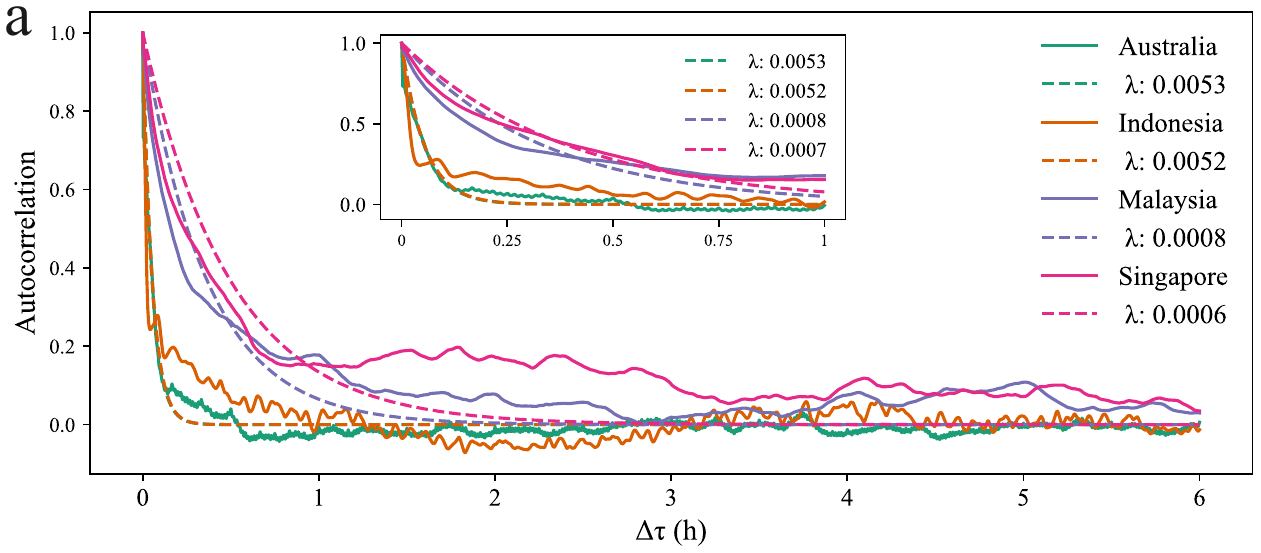}
  \label{fig6a}
\end{subfigure}
\vspace{-1\baselineskip}
\begin{subfigure}{\columnwidth}
  \centering
  \includegraphics[width=\linewidth]{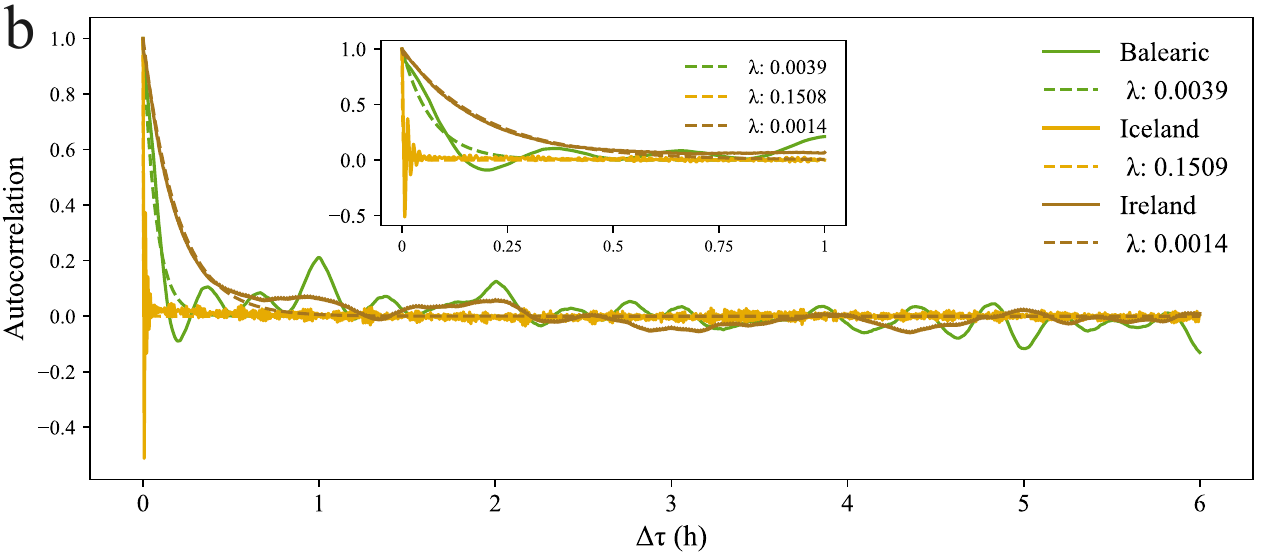}
  \label{fig6b}
\end{subfigure}

\captionsetup{justification=justified,singlelinecheck=false}
\caption{\textbf{Decay of the autocorrelation}. We calculate the autocorrelation for each region over a time lag of up to 6 hours. The solid lines represent the autocorrelation, while the dashed lines correspond to the exponential fit with a decay constant $\lambda$.}
\vspace*{-1.5em}
\label{fig6}
\end{figure}

As shown in Fig.~\ref{fig6}, the autocorrelation of these regions' power-grid frequencies exhibits an approximate exponential decay pattern concerning the time lag $\Delta \tau$.
To quantify the decay trend, we fit a curve with $e^{-\lambda \Delta \tau}$ (where $\lambda$ represents the decay constant and $\Delta \tau$ is the time lag) using an exponential model.
We observe that Iceland exhibits the highest decay constant value, measuring 0.1509, indicating a relatively rapid decay in autocorrelation.
On the other hand, Singapore shows the lowest decay value at 0.0006, suggesting long-lasting correlations, potentially arising from correlated noise. The results are robust regardless of whether we consider 1 or 6 hours of data for the exponential fit.

To quantify these emerging correlations, we calculate the Hurst exponent of power-grid frequency for each region.
Specifically, we estimate the Hurst exponent of these time series using the Detrended Fluctuation Analysis (DFA) method \cite{Peng1994, Peng1995, graves2017brief}.
Several studies have successfully applied DFA to analyze power-grid frequency data and uncover underlying long-range correlations \cite{gorjao2022phase,meyer2020identifying}.
DFA stands out among other methods for its ability to accurately quantify the strength of long-range correlations, even when dealing with non-stationary time series \cite{kantelhardt2001detecting}.

\begin{figure}
\centering
\begin{subfigure}{0.5\textwidth}
  \centering
  \includegraphics[width=\linewidth]{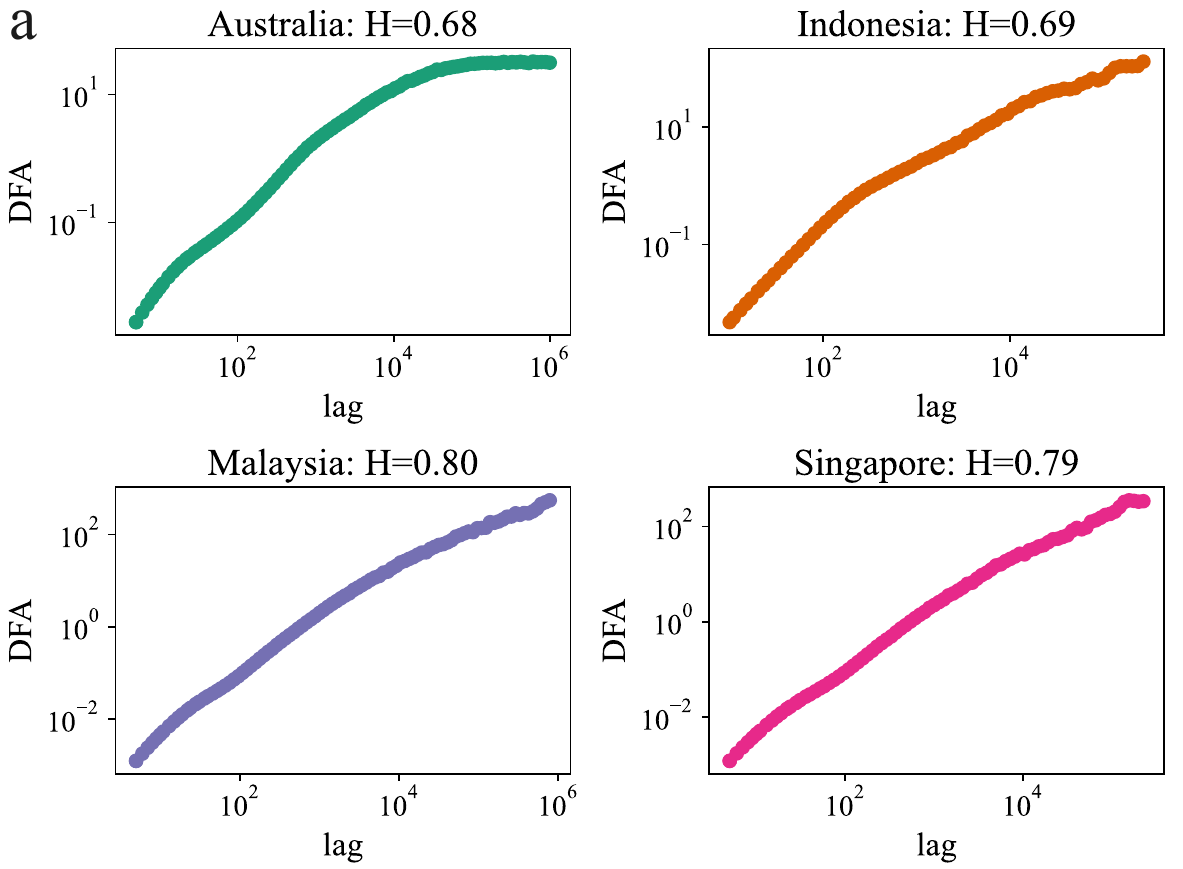}
  
  \label{fig7a}
\end{subfigure}\hfill
\begin{subfigure}{0.5\textwidth} % Specify the width for each subfigure
  \centering
  \includegraphics[width=\linewidth]{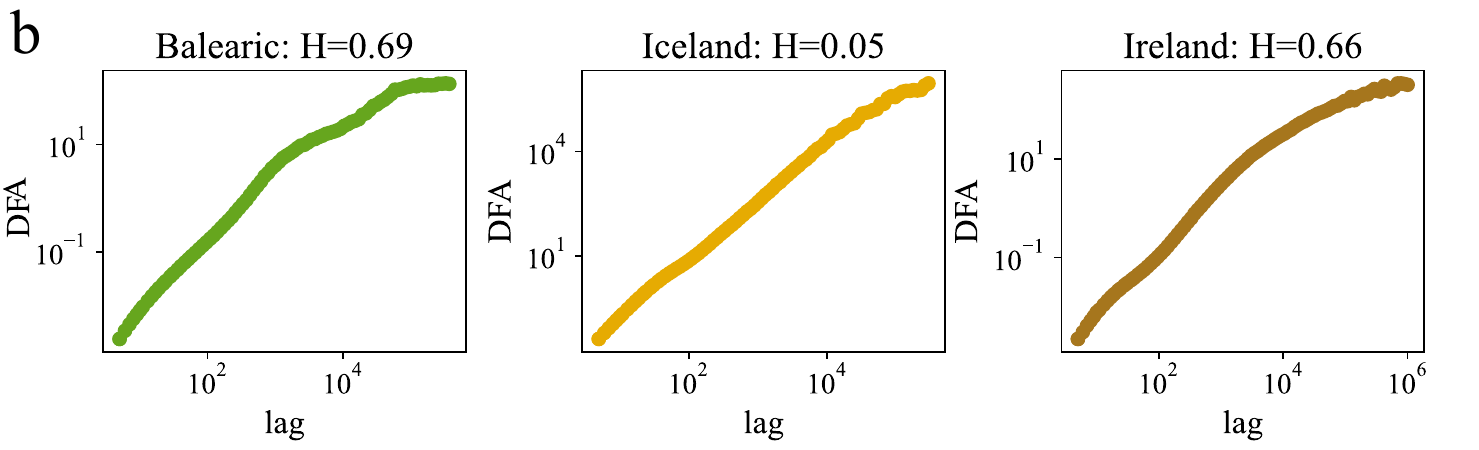}
  
  \label{fig7b}
\end{subfigure}
\captionsetup{justification=justified,singlelinecheck=false}
\caption{\textbf{Computing Hurst exponents}. We perform a Detrended Fluctuation Analysis. With the exception of Iceland, all regions exhibit Fluctuation Function patterns typified by Hurst values greater than 0.5.
}
\vspace*{-1.5em}
\label{fig7}
\end{figure}

We generate a set of lag values that span from 5 to $10^6$.  
Fig.~\ref{fig7} illustrates the DFA results of the power-grid frequency in various synchronous areas, utilizing the fluctuation function plotted against lag values. The slope of the fitted line in the log-log scale is equal to the Hurst index plus 1, in consideration of the integration performed in the DFA algorithm. From the slope of the Fluctuation Function, we extract the Hurst exponents, which exceed 0.5 for all regions except Iceland. This indicates the presence of positively correlated motions.

These observed correlations indicate that the stochastic dynamics $\xi$ in \eqref{eq:genericModel} might be better modeled as colored or fractional noise instead of a simple white noise process. Meanwhile, the almost exponential decay of the autocorrelation supports simple stochastic models, such as Ornstein-Uhlenbeck processes or extensions thereof.

\section{Discussion and conclusion}\label{sec:discussion}

%Brief overall summary
In the present study, we have collected and analyzed non-standard characteristics of the power-grid frequency data from Asia, Australia, and Europe. In particular, we demonstrated clear deviations from Gaussianity of both the frequency and its increment statistics, varying degrees of bimodality in different synchronous areas, and long-term correlations. All data are available openly on \href{https://power-grid-frequency.org/}{power-grid-frequency.org} \cite{power_grid_frequency} and our code to quantitatively compare models and data is available on Github \cite{github}. We advance previous data analysis \cite{anvari2020stochastic,rydin2020open} by including more non-European grids in our analysis. We compare the observed statistical results to reference systems and provide insights into the similarities and differences in power-grid dynamics across regions. 

%Frequency and frequency increment analysis + bimodality
Naively, we could expect that a large number of random perturbations on the power grid leads to Gaussian distributions. Instead, we observe very clear non-Gaussian distributions: Frequency statistics are highly bimodal and the frequency jumps (increments) are heavy-tailed. While the exact nature and origin of these properties might vary between different synchronous areas, we may speculate that a bimodal distribution could arise from deadbands in the control \cite{8626538,10202986} or transitions between two discrete states of the system. These transitions could be stochastic or deterministic, depending on the power system. Similarly, heavy tails in the increments are both explained plausibly as arising from sudden changes in power generation or load \cite{anvari2016short,anvari2020stochastic} as well as by deterministic changes due to power dispatch at the start of an hour \cite{schafer2018isolating,gorjao2020data}. 

%correlation Markov, MFDFA
As with Gaussianity, a common assumption about stochastic processes is to regard them as uncorrelated, i.e. as Markov processes. Again, we found that power grids both in Asia and Europe are more complex than this simple assumption. 
Furthermore, utilizing Detrended Fluctuation Analysis (DFA), we demonstrated that the Hurst values are generally greater than 0.5, i.e. that the time series displays a positive correlation. This finding is consistent with previous studies in this field \cite{schafer2023microscopic}. Power grids still exhibit a large number of synchronous generators thereby rotating mass with inertia. This already makes more continuous and correlated dynamics plausible. 
In addition, fluctuations both from the consumer \cite{anvari2022data} and from the generation side will often be correlated \cite{anvari2016short}. These results do not support a Markov property for all regions. Meanwhile, the full data displayed mostly linear properties, potentially simplifying at least one modeling aspect.

%Comparison between regions
Let us review the added benefit of considering measurements from diverse geographic regions, instead of limiting ourselves to synchronous areas from one continent.
%Comparing the data from different synchronous areas in different regions of the world leads to several insights: 
While Singapore displayed a pronounced bimodality, there is no clear trend that Asian synchronous areas are more bimodal than European ones. 
Synchronous areas in Asia, Australia, and Europe all displayed heavy tails and mostly linear dynamics, with the two most non-linear areas (Balearic and Singapore) from different geographic regions. 
The correlation results differed the most: The Asian data sets returned a slightly higher Hurst exponent, while Iceland is the only anti-correlated area in our data set.
Overall, we conclude that we require many different synchronous areas to have access to interesting dynamics as in Singapore or Iceland. Including data from multiple geographic regions will likely increase the chance of observing non-standard effects that have to be incorporated into any general-purpose model.

%key take-aways: What does that mean for the reader?
Concluding, our findings advance our understanding of power grids and their simulations. 
Regardless of the origin of the added complexity (bimodal, non-Gaussian fluctuations, non-Markov), power grid models, such as the simplified \eqref{eq:genericModel} should take these deviations into account and benchmark their models against empirical data, whether they are using linear \cite{cidras2002linear, kanna2015distributed,wang2016estimating} or non-linear \cite{8626538,10202986} models to be applicable to as many different settings as possible.

%Outlook/Future work
In the future, there are several other prospective studies and analytical options to continue the work presented here. 
One intriguing area of research is the comparison of frequency dynamics between different seasons, particularly winter and summer. 
Seasonal variations in power demand and generation patterns can significantly influence frequency behavior, and exploring these variations could provide valuable insights into the system's response to changing operational conditions.
Furthermore, investigating the interdependencies and interactions between power-grid frequency and other critical system variables, such as voltage amplitudes and power flows, could be beneficial to gain a more comprehensive understanding of the system's behavior.
Finally, our examination can be extended to islanded and microgrids operated primarily using power electronics to assess if different statistical and stochastic properties are present.

\section{Acknowledgments}\label{sec:Acknowledgments}
We gratefully acknowledge funding from the Helmholtz Association and the Networking Fund through Helmholtz AI and under grant no. VH-NG-1727, as well as the Scientific Research Projects Coordination Unit of Istanbul University, Project no. 39071. 
Map data copyrighted by OpenStreetMap contributors are available from \url{https://www.openstreetmap.org}.

\bibliographystyle{elsarticle-num}
\bibliography{Reference.bib}

% that's all folks
\end{document}